\documentstyle[epsfig,cesj]{article}

\year{2003}
\issue{1}
\rightmark{M. Rossi  and L. Zaninetti}
\leftmark{M. Rossi  and L. Zaninetti}
\title{
The cubic period-distance relation for
the  Kater reversible
pendulum
}
\author{M. Rossi     \inst{1}\email{E-mail: michele.rossi@unito.it},
        L. Zaninetti \inst{2}\email{E-mail: zaninetti@ph.unito.it}}
\institute{Dipartimento di Matematica, \\
Universit\`a degli Studi di Torino    \\
Via Carlo Alberto 10, 10123 Torino, Italy
\and    Dipartimento  di Fisica Generale,\\
Universit\`a degli Studi di Torino    \\
via P.Giuria 1,  10125 Torino,Italy}
\received{  }
\revised {  }
\abstract{
We describe the correct cubic relation between the mass
configuration of a Kater reversible pendulum and its period of
oscillation. From an analysis of its solutions we conclude that
there could be as many as three distinct mass configurations for
which the periods of small oscillations about the two pivots of the
pendulum have the same value.
We also discuss a real compound Kater
pendulum that realizes this property.
}

\begin{document}

\firstpage{1}
\maketitle \setcounter{page}{1}%

\section{Introduction}

A well known consequence of the fundamental equation of rotational dynamics
is that the period of small oscillations of a physical pendulum is given
by
\begin{equation}
T=\frac{2\pi }{\omega }=2\pi \sqrt{\frac{I}{mgh}}  \label{ph-period}
\end{equation}
where $m$ is total mass of the pendulum, $I$ its moment of inertia with
respect to the center of oscillation $O$ and $h$ the distance of the center
of mass from $O$. Then a physical pendulum oscillates like a simple
pendulum
of length
\begin{equation}
l=\frac{I}{mh}=\frac{gT^{2}}{4\pi ^{2}}  \label{eq.length}
\end{equation}
which is called the \textit{equivalent length }of our physical pendulum.

\noindent By the Huygens--Steiner theorem (also known as the ``parallel
axis
theorem'') it is possible to write
\[
I=mh^{2}+I_{0}
\]
where $I_{0}$ is the moment of inertia with respect to the center of mass.
By squaring equation (\ref{ph-period}) we get the following quadratic
relation
\begin{equation}
h^{2}-lh+\frac{I_{0}}{m}=0  \label{quadratic pdr}
\end{equation}
When $l^{2}-4I_{0}/m\geq 0$ that equation admits two real solutions $%
h_{1},h_{2}$ such that
\begin{equation}
h_{1}+h_{2}=l  \label{sum}
\end{equation}
In 1817 Captain H.Kater thought to use this last relation to empirically
check the Huygens--Steiner theorem. At this purpose he constructed
his
\textit{reversible pendulum} consisting of a plated steel bar equipped with
two weights, one of which can be moved along the bar.
This pendulum is
reversible because it can oscillate about two different
suspension points
realized by two knife edges symmetrically located on the bar.
By adjusting
the movable weight, it is possible to obtain a pendulum mass configuration
such that the periods about the two pivots coincide, the equivalent length
$%
l $ is the distance between the two knife edges and condition (\ref{sum})
is
satisfied.

The measurement of such a common period $T$, of the total mass $m$ and of
the distance $l$ between the two knife edges, gives then an easy way to
perform an empirical measurement of the earth's (apparent) gravitational
acceleration $g$ by applying formula (\ref{eq.length}). This is why the
Kater reversible pendulum is one of the favourite instrument to measure $g$
in student labs.

\noindent Anyway there is a subtle point in this procedure which is the
determination of the \textit{right mass configuration }of the pendulum.
This
problem gives rise to the following two questions:

\begin{enumerate}
\item  how many possible positions of the movable weight determine a
``good'' mass configuration for which the periods of small oscillations
about the two pivots coincide?

\item  when a good mass configuration is realized, is the equivalent
length $%
l$ necessarily represented by the distance between the pivots?
\end{enumerate}

\noindent If the answer to the second question is assumed to be ``yes''
then
the quadratic equation (\ref{quadratic pdr}) gives precisely \textit{two }%
possible good mass configurations since $h$ depends linearly on the
position
$x$ of the movable weight. These mass configurations can then be
empirically
obtained by the following standard procedure \cite{peters}:

\begin{itemize}
\item  by varying the movable mass position $x$ collect two series of data
$%
\left( x,T\right) $, one for each pivot,

\item  make a parabolic fitting of the data by means of two parabolas of
the
following type:
\begin{equation}
T=ax^{2}+bx+c  \label{parabola}
\end{equation}

\item  these parabolas meet in at most two points $\left( x_{1},T\right)
$, $%
\left( x_{2},T\right) $: positions $x_{1}$ and $x_{2}$ determine the two
desired good mass configurations.
\end{itemize}

\noindent Such a parabolic fitting is justified by two considerations. The
first one is that we are looking for \textit{two }good mass configurations,
then the fitting curves have to admit at most two intersection points. The
second one is the empiric observation of the data which apparently seem to
be arranged just along two convex parabolas with vertical axis.

This is what is usually done although \textit{the right answer to the
second
question should be ``no''}, as was firstly pointed by
Shedd and
Birchby in 1907 \cite{shedd1,shedd2,shedd3}.
Their remark seems to
have escaped general
attention, perhaps due to the fact that, if the pendulum is well assembled,
the previous parabolas meet at points whose abscissas give \textit{almost
exactly} the good mass configurations having the distance between pivots as
equivalent length. The latter is much easier determined than any other
equivalent length associated with further good mass configurations of the
pendulum \cite{candela}! But what does mean ``well assembled''?

\begin{figure}[htbp]
\includegraphics[width=10cm,angle=0]{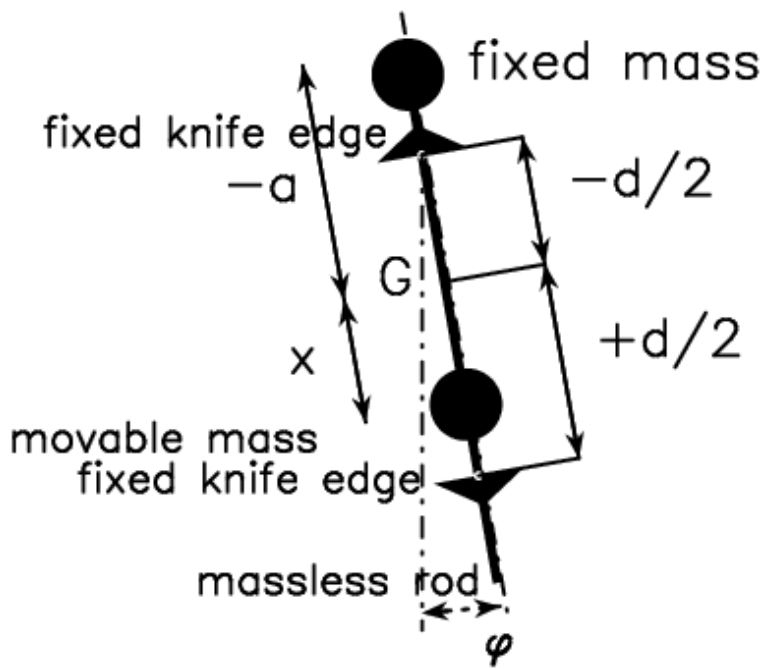}
\caption{\label{fig_01}
Front view of idealized
pendulum with two point masses and
a massless rigid rod.
}
\end{figure}

To fix ideas consider an ``ideal'' Kater pendulum consisting of an
idealised
massless rigid rod ($x$-axis) supporting two identical point masses, $m_{f}
$
fixed at $-a$ and $m_{m}$ at a variable position $x$. The assembly has two
distinct suspension points for the oscillations positioned at $-d/2$ and
at $%
+d/2$
%inizio proposta
, a sketch of this "ideal" pendulum is reported
in Figure~\ref{fig_01}.
%figure fig_01
%fine proposta
 Then $d$ is the distance between the pivots, the center of mass is
located at
\[
b=\frac{-am_{f}+xm_{m}}{m_{f}+m_{m}}=\frac{x-a}{2}
\]
%inizio proposta
and  the moment of inertia about the center of mass at b
is given by
%fine proposta
\[
I_{0}=\left( b+a\right) ^{2}m_{f}+\left( b-x\right) ^{2}m_{m}=\frac{m}{4}%
\left( x+a\right) ^{2}
\]
where $m=2m_{f}=2m_{m}$ is the total mass of the pendulum.
%inizio proposta
The moment of inertia   $I_{1}$
and $I_{2}$  with respect to the two pivots are
\begin{equation}
I_{1}=(\frac {a+x} {2})^2  m + (\frac{d}{2} +b)^2 m
\quad ,
\nonumber
\end {equation}
and
\begin{equation}
I_{2}=(\frac {a+x} {2})^2  m + (\frac{d}{2} -b)^2 m
\quad .
\nonumber
\end {equation}
%fine proposta
When $x$
determines a good mass configuration the resulting periods $T_{1}$ and $%
T_{2} $ of small oscillations about the two pivots, respectively, have
equal
values. Equation (\ref{ph-period}) gives
\begin{eqnarray*}
T_{1} &=&2\pi \sqrt{\frac{m(b+\frac{d}{2})^{2}+\frac{m}{4}\left( x+a\right)
^{2}}{mg\left| b+\frac{d}{2}\right| }}=2\pi \sqrt{\frac{\left( x-a+d\right)
^{2}+\left( x+a\right) ^{2}}{2g\left| x-a+d\right| }} \\
T_{2} &=&2\pi \sqrt{\frac{m(b-\frac{d}{2})^{2}+\frac{m}{4}\left( x+a\right)
^{2}}{mg\left| b-\frac{d}{2}\right| }}=2\pi \sqrt{\frac{\left( x-a-
d\right)
^{2}+\left( x+a\right) ^{2}}{2g\left| x-a-d\right| }}
\end{eqnarray*}
Then $T_{1}^{2}=T_{2}^{2}$ gives a cubic equation on the variable $x$. If
it is assumed that
\begin{equation}
\left( x-a \right)^2 - d^2 < 0 \label{assump.}
\end{equation}
which occurs, for instance, when suspension points are the end points of
the pendulum bar, one finds that
\begin{equation}
\left( x-a\right) \left[ 2\left( x^{2}+a^{2}\right) -d^{2}\right] =0
\label{ideal:eq}
\end{equation}
Its solutions are then given by
%\begin{subequations}
\begin{eqnarray}
\label{xsol}
x&=&a,\label{xsol.a}\\
x&=&\pm \sqrt{\frac{d^{2}}{2}-a^{2}} \label{xsol.b}
\,,
\end{eqnarray}
%\end{subequations}
which represent \textit{all the possible positions of the movable weight
giving a good mass configuration for the ideal Kater pendulum}. The first
solution, $x=a$, always exists. Furthermore, if $d/\sqrt{2}\geq a$ there
are
two additional positions which are symmetric with respect to the origin
i.e.
the middle point of the massless bar. Recall formula (\ref{eq.length}) to
obtain the associated equivalent lengths. For the last two symmetric
solutions it gives
\[
l=d
\]
But the equivalent length associated with the first solution is
\begin{equation}
l^{\prime }=\frac{d}{2}+2\frac{a^{2}}{d}
\label{lidealized}
\end{equation}
which in general does not coincide with the distance $d$ between the two
pivots.

\noindent On the other hand if (\ref{assump.}) is not assumed and we are
in the more ``pathological'' case of a pendulum such that $\left( x-a
\right)^2 - d^2 > 0$ then $T_1^2=T_2^2$ reduces to a linear equation in
the variable $x$ whose solution is
\[
x=-\frac{d^2}{4a}
\]
and the associated equivalent length is
\[
l^{\prime \prime}= a+\frac{d^2}{4a}
\]
which in general does not coincide with the distance $d$ between the two
pivots.

\noindent Therefore for an ideal Kater pendulum the answers to the previous
questions are:

\begin{enumerate}
\item  there are at most \textit{three }possible positions of the movable
weight which determine a good mass configuration;

\item  no; there always exists a good mass configuration whose associated
equivalent length does not coincide with the distance between pivots.
\end{enumerate}

\noindent An immediate consequence is that a parabolic fitting of the
empirical data $\left( x,T\right) $ can't be the best fitting since two
parabolas never meet at three points! Moreover in some particular cases a
parabolic fitting may cause strong distortions in determining good mass
configurations. For example:

\begin{itemize}
\item  if either $d/\sqrt{2}<a$ or (\ref{assump.}) is not satisfied, the
ideal Kater pendulum admits a unique good
mass configuration; typically a parabolic fitting of data in
this situation gives parabolas meeting only at imaginary points and the
procedure stops;

\item  if $a\sim \pm d/2$ then the first solution of (\ref{ideal:eq}) is
quite near to one of the two further symmetric solutions; a parabolic
fitting of data gives only two intersection points but we do not know if
one
(and which one?) of them is nearer to the position associated with $l$ than
to the one associated with $l^{\prime }$; in this situation also $l\sim
l^{\prime }$ but they are not equal; then associating $l$ with a so
determined good mass configuration may cause a relevant error in the final
value of $g$.
\end{itemize}

\noindent One may object that we are discussing an empiric procedure by
means of an ideal pendulum. In particular the position $x=a$ for the
movable
mass gives the completely symmetric mass configuration with respect to the
middle point of the ideal pendulum bar. When a physical pendulum with $%
m_{f}\neq m_{m}$ is considered, what is such a mass configuration? Does it
occur again?

The answer is ``yes''. The key observation is that, for both pivots, the
variable position $x$ of the movable mass and the resulting period $T$ of
small oscillations are related by a cubic expressions of the following type
(period-distance relations)
\begin{equation}
ax^{2}+bx+c=T^{2}+dxT^{2}\,,  \label{cubic pdr}
\end{equation}
where the coefficients $a,b,c,d$ depend on the pendulum parameters. This is
precisely what Shedd and Birchby pointed out in their papers
\cite{shedd1,shedd2,shedd3}  giving
theoretical and empirical evidences: they called the two (one for each
pivot) equations (\ref{cubic pdr}) \textit{the equations of the reversible
pendulum} (see equations (10) and (11) of their first paper). Here we will
refer to (\ref{cubic pdr}) as \textit{the cubic period--distance
relation }%
of the physical Kater pendulum considered. Note that only coefficients $%
a,b,c,d$ depend on the pendulum parameters, while the polynomial type of
equation (\ref{cubic pdr}) does not depend on the choice of the pendulum.
Thus, we can reduce the search for good mass configurations to a simple
cubic equation similar to Eq.~(\ref{ideal:eq}).

A first point in the present paper is to give a mathematically rigorous
proof of the following

\begin{theorem}
\label{Th}\textit{Let }$p_{1}\left( x,y\right) ,p_{2}\left( x,y\right) $%
\textit{\ be the following cubic polynomials}
\[
p_{i}\left( x,y\right) =A_{i}x^{2}+B_{i}x+C_{i}-y^{2}-D_{i}xy^{2}\quad
,\quad i=1,2\ .
\]
\textit{where }$A_{i},B_{i},C_{i},D_{i}$\textit{\ are real coefficients
and $%
D_{1}\neq D_{2}$ . Then they admit always two real common roots and two
pairs of complex conjugated common roots which may be real under suitable
conditions on coefficients} $A_{i},B_{i},C_{i},D_{i}$. \textit{Thinking
them
as points in the complex plane }$\left( x,y\right) $ \textit{they are
symmetric three by three with respect to the} $x$\textit{--axis}. \textit{%
Moreover these are all the common roots they can admit (that is: all the
further common roots are ``at infinity'').}
\end{theorem}

\noindent This algebraic result leads to the following physical statement:

\begin{corollary}
\label{PS}\textit{A physical Kater pendulum, with a ``sufficiently long''
bar, admits always a ``good'' mass configuration whose associated
equivalent
length does not in general coincide with the distance between the pivots.}

\textit{Under ``suitable conditions'' on the pendulum parameters, it may
admit two further ``good'' mass configurations. They correspond to
symmetric
positions of the movable mass, with respect to the middle point of the bar
(if also the pivots are symmetrically located). They admit a common
associated equivalent length which is precisely the distance between the
pivots.}

\textit{Moreover the pendulum can't admit any further good mass
configuration.}
\end{corollary}

\noindent We will specify the meaning to the vague expressions
``sufficiently long'' and ``suitable conditions''.

Although Shedd and Birchby knew in practice the content of the previous
statement (they actually wrote down all the three good mass configurations
in period--distance terms -- see formulas (27) of their first paper) they
couldn't give a rigorous proof of it. They studied the geometry of the
curves determined by the cubic period--distance relations by means of an
old
and non--standard ``Newton's classification''. They then arrived to
conclude
that (see the bottom lines of p. 281 in their first paper):

\begin{description}
\item  \textit{``Of the nine possible intersections of two cubic curves, in
the present case three are imaginary or at infinity, three belong to the
condition that }$T$\textit{\ is negative, and three belong to positive
values of }$T$\textit{, and can hence be experimentally realized.''}
\end{description}

\noindent This conclusion does not exclude that, under some suitable
conditions on the pendulum parameters, at least two of the three
``imaginary
or at infinity'' intersections may become real and maybe physically
realizable giving more than three good mass configurations. Actually we
will
see that these three intersections are not imaginary but definitely ``at
infinity'' and they can never give physical results.

A second aim of the present paper is to observe that \textit{the best
fitting of empirical data }$\left( x,T\right) $ \textit{is then given by
two
cubic curves of type }(\ref{cubic pdr}) instead of two parabolas of type (%
\ref{parabola}). We will support this remark by experimental evidence for a
real compound Kater pendulum.

The paper is organised as follows. Section~\ref{sec2} is devoted to prove
Theorem \ref{Th} and the physical statement of Corollary \ref{PS}. Here
we set the main notation and describe the physics of a real Kater pendulum.
The proof of Theorem \ref{Th} is based on elementary elements of complex
algebraic and projective geometry. Anyway a non--interested reader may skip
it without losing any useful element to understand what follows. In
Sec.~\ref
{sec3} we describe an effective experiment. Section~\ref{sec4} is devoted
to
the analysis of experimental data by a linear fit of the period-distance
cubics. An estimate of their intersection points gives the good mass
configurations and the associated periods for our real Kater pendulum. Then
the value of $g$. A comparison with a parabolic fitting of data is then
given.

Appendix \ref{appa} is devoted to discuss the ``suitable conditions'' on
the
pendulum parameters under which the pendulum admits all the possible good
mass configurations (see Corollary \ref{PS}). In Appendix~\ref{appc} we
collect some further numerical methods to analyse our empirical data.

\section{Physics of the Kater reversible pendulum}

\label{sec2}

%figure fig_02
\begin{figure}[htbp]
\begin{center}
\includegraphics[width=10cm,angle=0]{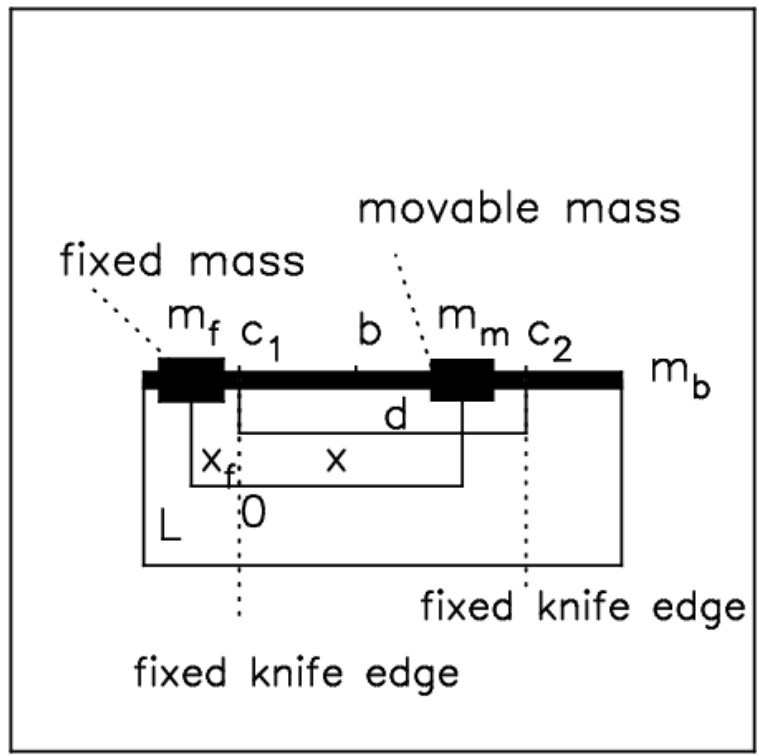}
\end{center}
\caption{\label{fig_02}Detailed side view of the Kater pendulum
(not to scale).}
\end{figure}

\begin{figure}[htbp]
\begin{center}
\includegraphics[width=10cm,angle=0]{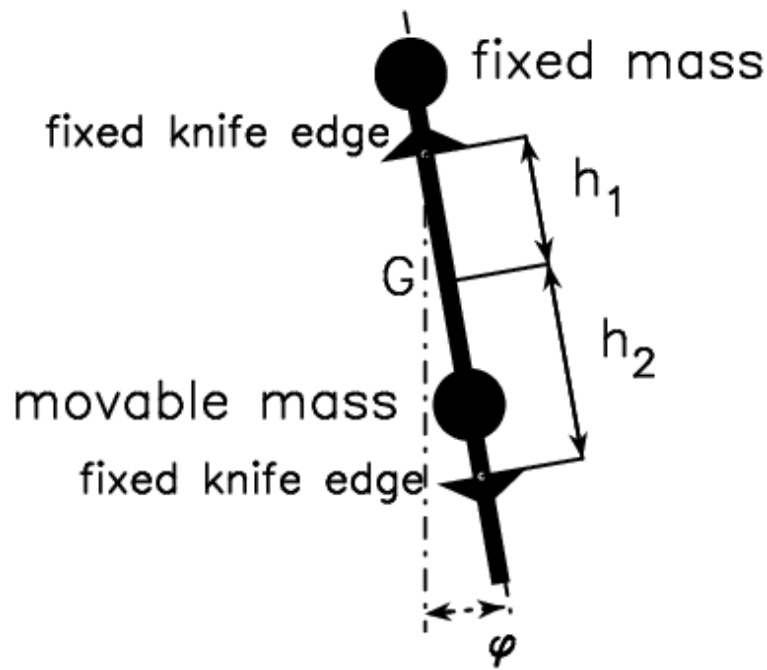}
\end{center}
\caption{\label{fig_03}Front view of the Kater pendulum (not to
scale). The pendulum swings in the plane of the picture; its
pivot can be inverted.}
\end{figure}

\textit{Notation. }Consider a physical Kater pendulum composed of a rigid
bar equipped with two weights (see Fig.~\ref{fig_02} and Fig.~\ref
{fig_03}). The pendulum can be suspended by two knife--edges, $c_{1}$
and $%
c_{2}$, symmetrically located on the bar. The weight $m_{f}$ is placed in a
fixed position which is not between the knives. The other one, $m_{m}$, can
be moved along the bar. Small oscillations of the pendulum are
parameterised
by an angle $\varphi $ such that $\varphi \approx \sin \varphi $, that is,
$%
\varphi ^{3}\approx 0$. The equation of motion of the pendulum is then
given
by
\begin{equation}
\ddot{\varphi}+\frac{mgh_{i}}{I_{i}}\varphi =0\,,  \label{eq.moto}
\end{equation}
where $g$ is the earth's apparent gravitational acceleration, $m$ is the
total mass of the pendulum, $h_{i}$ is the distance of the center of mass
from the knife--edge $c_{i}$, and $I_{i}$ is the moment of inertia about $%
c_{i}$.

%figure fig_03

The Steiner's theorem~\cite{resnick} asserts that
\begin{equation}
I_{i}=I_{0}+mh_{i}^{2}\,,
\end{equation}
where $I_{0}$ is the moment of inertia with respect to the center of mass.
The associated period of small oscillations is
\begin{equation}
T_{i}=\frac{2\pi }{{\omega }_{i}}=2\pi \sqrt{\frac{I_{i}}{mgh_{i}}}=2\pi
\sqrt{\frac{I_{0}+mh_{i}^{2}}{mgh_{i}}}\,.  \label{equation:t}
\end{equation}
Equation~(\ref{equation:t}) implies that the Kater pendulum oscillates with
the same period as a simple pendulum whose length is given by
\begin{equation}
l_{i}=\frac{I_{i}}{mh_{i}}=\frac{I_{0}+mh_{i}^{2}}{mh_{i}}\,.  \label{li}
\end{equation}

Assume now that the movable mass $m_{m}$ is placed at a point $x_{0}$ on
the
bar such that
\begin{equation}
T_{1}=T_{2}=T(x_{0})\,.  \label{equation:T1=T2}
\end{equation}
Such a point will be called \textit{a characteristic position of the
pendulum. }Equation~(\ref{equation:T1=T2}) can be satisfied if and only if
$%
l_{1}=l_{2}=l(x_{0})$. The length $l=l(x_{0})$ will be called \textit{the
characteristic length of the pendulum associated with the characteristic
position} $x_{0}$. Analogously the associated periods $T(x_{0_{j}})$ will
be
\textit{the characteristic periods of the pendulum}. The knowledge of $%
l(x_{0_{j}})$ and $T(x_{0_{j}})$ for each $j=1,2,3$ yields the value of $g$
from the relation
\begin{equation}
T=2\pi \sqrt{\frac{l}{g}}\,,  \label{period}
\end{equation}
and therefore
\begin{equation}
g=\frac{4\pi ^{2}}{T^{2}}\,.  \label{g}
\end{equation}
The variable position of $m_{m}$ is described by a linear coordinate $x$
having origin at $c_{1}$. Then $c_{2}$ is the point $x=d>0$ (see Fig.~\ref
{fig_02}) while the fixed weight is placed at $x_{f}$ such that $%
(d-L)/2<x_{f}<0$.

\noindent The movable and fixed weights are composed of disks whose radii
are given respectively by $r_{m}$ and $r_{f}$.

\noindent $L$ is the length of the pendulum bar.

Therefore,
%inizio proposta
the distance $h$ between the pendulum center of mass
%fine proposta
and the
origin $c_{1}$ depends on the position $x$ of $m_{m}$ and is given by:
\begin{equation}
h=\frac{\frac{d}{2}m_{b}+x_{f}m_{f}+xm_{m}}{m_{b}+m_{f}+m_{m}}\,,
\end{equation}
where $m_{b}$ is the mass of the bar. Set
\begin{equation}
m=m_{b}+m_{f}+m_{m}\,,
\end{equation}
and
\begin{equation}
K=\frac{\frac{d}{2}m_{b}+x_{f}m_{f}}{m}\,.
\end{equation}
Then $h$ can be rewrite as follows
\begin{equation}
h=K+\frac{m_{m}}{m}x\,.  \label{h}
\end{equation}
The moment of inertia $I_{0}$ is then given by
\begin{equation}
I_{0}=\left( h-x_{f}\right) ^{2}m_{f}+\left( h-x\right) ^{2}m_{m}+\left( h-
\frac{d}{2}\right) ^{2}m_{b}+I_{0}^{\prime \prime }\,,  \label{I0}
\end{equation}
where
\begin{equation}
I_{0}^{\prime \prime }=\frac{r_{f}^{2}}{2}m_{f}+\frac{r_{m}^{2}}{2}m_{m}+%
\frac{L^{2}}{12}m_{b}\,.  \label{I''0}
\end{equation}
Set
\begin{equation}
I_{0}^{\prime }=I_{0}^{\prime \prime }+m_{f}\left( x_{f}-K\right)
^{2}+m_{b}\left( \frac{d}{2}-K\right) ^{2}+m_{m}K^{2}\,,  \label{I'0}
\end{equation}
and $I_{0}$ can be rewritten as follows:
\begin{equation}
I_{0}=m_{m}\frac{m-m_{m}}{m}x^{2}-2m_{m}Kx+I_{0}^{\prime }\,.  \label
{I0bis}
\end{equation}
From Eq.~(\ref{li}) the condition (\ref{equation:T1=T2}) is satisfied if
and
only if
\begin{equation}
\frac{I_{0}+mh_{1}^{2}}{mh_{1}}=\frac{I_{0}+mh_{2}^{2}}{mh_{2}}\,,
\label{equality}
\end{equation}
which is equivalent to requiring that
\begin{equation}
\left( h_{1}-h_{2}\right) \left( mh_{1}h_{2}-I_{0}\right) =0\,.
\label{cubiceq2}
\end{equation}
From Eq.~(\ref{h}) we have
%\begin{subequations}
\begin{eqnarray}
\label{hi}
h_1 & = & h = K + \frac{m_m}{m} x \\
h_2 &
= & d - h = d - K - \frac{m_m}{m} x
\,,
\end{eqnarray}
%\end{subequations}
and we get the first characteristic position by imposing that $h_{1}=h_{2}
$,
that is,
\begin{equation}
x_{0_{1}}=\frac{d}{2}+\frac{m_{f}}{2m_{m}}\left( d-2x_{f}\right) \,.
\label{char.pos.1}
\end{equation}
Two additional characteristic positions can be obtained by the second
factor
in Eq.~(\ref{cubiceq2}). By letting $mh_{1}h_{2}-I_{0}=0$ and expressing $%
I_{0}$ as in Eq.~(\ref{I0bis}), we have
\begin{equation}
x^{2}-dx-\frac{mK^{2}-mdK+I_{0}^{\prime }}{m_{m}}=0\,,
\end{equation}
whose solutions are
%\begin{subequations}
\begin{eqnarray}
\label{char.pos.2,3}
x_{0_2} & = & \frac{d}{2}+ \frac{1}{2}\sqrt{d^2 + 4
\frac{m K^2 - m d K + I_0^{\prime}}{m_m}} \\ x_{0_3} &
= &
\frac{d}{2} - \frac{1}{2}\sqrt{d^2 + 4 \frac{m K^2 - m d K +
I_0^{\prime}}{m_m}}
\,.
\end{eqnarray}
%\end{subequations}
To determine the associated characteristic lengths $l(x_{0_{j}})$ use Eqs.~
(%
\ref{li}), (\ref{equality}), and (\ref{hi}). It follows that
\begin{equation}
l(x_{0_{j}})=\frac{I_{0}+m\left( K+\frac{m_{m}}{m}x_{0_{j}}\right) ^{2}}{%
mK+m_{m}x_{0_{j}}}\,.
\end{equation}
It is then easy to observe that $l(x_{0_{2}})$ and $l(x_{0_{3}})$ are equal
and constant because $x_{0_{2}}$ and $x_{0_{3}}$ are symmetric. Precisely
\begin{equation}
l(x_{0_{2}})=l(x_{0_{3}})=h_{1}+h_{2}=d\,,  \label{l2,3}
\end{equation}
and they do not depend on the other physical parameters of the pendulum. On
the contrary, this is not true for $l(x_{0_{1}})$ because
\begin{equation}
l(x_{0_{1}})=\frac{d}{2}+2\frac{I_{0}^{\prime \prime }}{md}+\frac{%
m_{f}\left( m_{m}+m_{f}\right) \left( d-2x_{f}\right) ^{2}}{2m_{m}md}\,.
\label{l1}
\end{equation}
The reader may compare the characteristic positions (\ref{char.pos.1}),
(\ref
{char.pos.2,3}) and the associated characteristic lengths (\ref{l1}), (\ref
{l2,3}), now obtained, with those given in Eq. (27) by Shedd and
Birchby\cite{shedd1}.

Moreover the \textit{period--distance relations }of the pendulum (what
Shedd
and Birchby called ``the equations of the Kater pendulum''
\cite{shedd1})
can be
obtained by Eq. (\ref{equality}) when $h_{1}$ and $h_{2}$ are expressed as
in Eqs. (\ref{hi}). When the pendulum oscillates about the pivot $c_{i}$,
the period $T_{i}$ and the distance $x$ results to be related by the
following cubic relations
\begin{equation}
A_{i}x^{2}+B_{i}x+C_{i}=T_{i}^{2}+D_{i}xT_{i}^{2}\quad ,\quad i=1,2\ ,
\label{equation:Ci}
\end{equation}
where
%\begin{subequations}
\begin{eqnarray}
\label{coeff1}
A_1 & = & \frac {4\pi^2 m_m} {g m K} \\
B_1 & = & 0 \nonumber \\ C_1& = & \frac {4 \pi^2} {g m
K} \left(I_{0}^{\prime} + m K^2 \right) \\ D_1& = & \frac
{m_m} {m K} \,,
\end{eqnarray}
%\end{subequations}
and
%\begin{subequations}
\begin{eqnarray}
\label{coeff2}
A_2& = & \frac {4\pi^2 m_m} {g m (d-K)}
\\ B_2 & = & - \frac {8 \pi^2 m_m d} {g m \left(
d-K\right)} \\ C_2 & = & \frac {4 \pi^2} {g
m(d- K)} \left(I_{0}^{\prime} + m \left( d - K \right)^2 \right)
\\ D_2 & = & - \frac {m_m} {m (d-K)}
\,.
\end{eqnarray}
%\end{subequations}
All the possible characteristic positions are then given by all the common
roots of Eqs. (\ref{equation:Ci}).\medskip

\textit{Proof of Theorem \ref{Th}.} For more details on the mathematics
here
involved see, for instance, Harris\cite{harris} or Shafarevich
\cite{shafarevich} among other introductory textbooks on algebraic geometry.

Consider $(x,y)$ as coordinates of points in the complex affine plane $%
\mathbf{C}^{2}$. Then equations $p_{1}\left( x,y\right) =0$ and $p_{2}\left
(
x,y\right) =0$ give two cubic complex algebraic curves, ${\mathcal C}_{1}$
and ${\mathcal C}_{2}$, whose intersection points are precisely the common
roots of $p_{1}$ and $p_{2}$. We can compactify $\mathbf{C}^{2}$ by
``adding
a line at infinity'': this procedure produces the complex projective plane
$%
\mathbf{P}_{\mathbf{C}}^{2}$. More precisely, we can consider our complex
variables $x$ and $y$ to be a ratio of further variables, that is,
\begin{equation}
x=\frac{X}{Z}\ \quad {and}\quad  y=\frac{Y}{Z}\,.
\end{equation}
The equations defining ${\mathcal C}_{1}$ and ${\mathcal C}_{2}$ multiplied
by
$Z^{3}$ become the following
%\begin{subequations}
\begin{eqnarray}
\label{proj.cubics}
A_1X^2 Z + B_1X Z^2 +C_1 Z^3 = Y^2 Z + D_1X Y^2
\\
A_2 X^2 Z + B_2 X Z^2 +C_2 Z^3 = Y^2 Z + D_2 X Y^2\,.
\end{eqnarray}
%\end{subequations}
which are the defining equations of the projective completions $\widetilde
{%
{\mathcal C}}_{1}$ and $\widetilde{{\mathcal C}}_{2}$, respectively. The main
ingredient of the present proof is the following

\begin{theorem}
\textbf{(Bezout)} \textit{Given two distinct irreducible complex algebraic
plane curves of degree }$d_{1}$ \textit{and }$d_{2}$\textit{, their
projective completions admits a finite number of intersection points.
Precisely if every intersection point is counted with its algebraic
multiplicity then this number is }$d_{1}d_{2}$\textit{.}
\end{theorem}

In particular the projective completions $\widetilde{{\mathcal C}}_{1}$ and
$%
\widetilde{{\mathcal C}}_{2}$ meet in 9 points, counted with their algebraic
multiplicities. The Bezout theorem is a consequence of the Fundamental
Theorem of Algebra which asserts that on the field $\mathbf{C}$ of complex
numbers every polynomial admits as many roots as its degree.

\noindent The first step is to study the intersections ``at infinity'',
that
is, which belong to the added ``line at infinity.'' The equation of this
line is $Z=0$ and by Eq.~(\ref{proj.cubics}) it intersects both our cubics
at $y_{\infty }$ (that is, the point $X=Z=0$ which is the infinity point of
the affine $y$-axis $x=0$) and in $x_{\infty }$ (that is, the point $Y=Z=0$
which is the infinity point of the affine $x$-axis $y=0$). Both of these
are
inflection points for ${\mathcal C}_{1}$ and ${\mathcal C}_{2}$. In $y_
{\infty
}$ the inflection tangent line of ${\mathcal C}_{1}$ is given by
\[
t_{1}:D_{1}X+Z=0\,,
\]
while the inflection tangent line of ${\mathcal C}_{2}$ is
\[
t_{2}:D_{2}X+Z=0\,.
\]
They cannot coincide since $D_{1}\neq D_{2}$. Therefore $y_{\infty }$ is a
simple intersection point of our cubics, that is, it admits intersection
multiplicity 1. On the other hand, in $x_{\infty }$ both ${\mathcal C}_{1}$
and ${\mathcal C}_{2}$ have the same inflection tangent line which is the
infinity line $Z=0$. Then $x_{\infty }$ has intersection multiplicity 2.
Consequently these infinity points count 3 of the 9 intersection points.
The
remaining 6 intersections must be affine, that is, they cannot belong to
the
compactifying line at infinity.

\noindent To find them note that for $i=1,2$
\begin{equation}
t_{i}\cap {\mathcal C}_{i}=y_{\infty }\,,
\end{equation}
with intersection multiplicity 3 because it is an inflection point for $%
{\mathcal C}_{i}$ with tangent line $t_{i}$. On the other hand
%\begin{subequations}
\begin{eqnarray}
t_1 \cap {\mathcal C}_2 & = & \{y_{\infty}, P_1,P_2\}
\\ t_2 \cap {\mathcal C}_1 & = & \{y_{\infty}, Q_1,Q_2\}
\end{eqnarray}
%\end{subequations}
where $P_{h}\not{=}y_{\infty }$, $Q_{k}\not{=}y_{\infty }$, and $P_{h}\not%
{=}Q_{k}$, because $t_{1}$ and $t_{2}$ are always distinct. Therefore, the
affine intersection points of ${\mathcal C}_{1}$ and ${\mathcal C}_{2}$
cannot
belong to the lines $t_{1}$ and $t_{2}$, and they can be recovered by
studying the common solutions to the following equations
%\begin{subequations}
\begin{eqnarray}
\label{system}
y^2 & = & \frac {A_1x^2 +B_1x +C_1}{1 + D_1 x} \\
y^2 & = & \frac {A_2x^2 + B_2x +C_2}{1 + D_2 x}
\,,\end{eqnarray}
%\end{subequations}
because those points do not make the denominators vanish. So they are
reduced to the roots of the following cubic equation
\begin{equation}
(A_{1}x^{2}+B_{1}x+C_{1})(1+D_{2}x)=(A_{2}x^{2}+B_{2}x+C_{2})(1+D_{1}x)\,.
\label{cub.eq.}
\end{equation}
It is a cubic equation with real coefficients. Therefore it admits 3
complex
roots one of which is surely a real number. The remaining two roots are
necessarily complex conjugated: their reality depends on the coefficients $
A_{i},B_{i},C_{i},D_{i}$.\medskip

\textit{Proof of Corollary \ref{PS}.} Recall the cubic period--distance
relations (\ref{equation:Ci}). Setting $T_{1}=T_{2}=y$ they are represented
by the two cubic curves ${\mathcal C}_{1}$ and $\mathcal{C}_{2}$ whose
coefficients are assigned by formulas (\ref{coeff1}) and (\ref{coeff2}),
respectively. Note that they are real numbers and $D_{1}\neq D_{2}$ since $
d\neq 0$. The hypothesis of Theorem \ref{Th} are then satisfied and the
characteristic positions of the pendulum must be represented by the \textit
{%
real affine intersection points admitting }$y\geq 0$\textit{.} To conclude
the proof observe that Eq. (\ref{cub.eq.}) divided out by $4\pi
^{2}/gmK\left( d-K\right) $ gives exactly the cubic equation (\ref
{cubiceq2}%
). The real root is then given by (\ref{char.pos.1}) and it always occurs
when
\[
\frac{m_{f}}{m_{m}}\left| d-2x_{f}\right| \leq L
\]
The remaining two roots are then assigned by (\ref{char.pos.2,3}). A
discussion of their reality is given in Appendix \ref{appa}.

\section{The experiment}

\label{sec3} The physical parameters characterising our pendulum are given
in Table~\ref{parameters}; the digits in parentheses indicate the
uncertainties in the last digit.
\begin{center}
\begin{table}
\caption{\label{parameters}The physical parameters characterising
the pendulum.}
\begin{tabular}{cccccccc}
\hline
$m_m$\,(g) & $m_f$\,(g) & $m_b$\,(g) & $x_f$\,(cm) &
$l$\,(cm) & $d$\,(cm) & $r_f$\,(cm) & $r_m$\,(cm) \\
\hline
1399(1) & 1006(1) & 1249(1) & $-26.73(1)$ &
167.0(1) & 99.3(1) & 5.11(1) & 5.12(1) \\
\hline
\end{tabular}
\end{table}
\end{center}

The bar length is measured by means of a
ruler whose accuracy is $\pm 1$\thinspace mm. The radii $r_{m}$ and $r_{f}$
and the position $x_{f}$ are measured by a Vernier caliper accurate to $\pm
0.1$\thinspace mm. The masses $m_{b}$, $m_{m}$, and $m_{f}$ are determined
by means of a precision balance accurate to one gram. With reference to the
structural conditions in Appendix~\ref{appa}, we are in the case 3.b i.e.
all the three characteristic positions occurs and precisely $%
x_{0_{2}},x_{0_{3}}$ are placed between the knives while $x_{0_{1}}$ is on
the opposite side of the bar with respect to $m_{f}$. By recalling Eqs.~
(\ref
{char.pos.1}) and (\ref{char.pos.2,3}), we expect that
%\begin{subequations}
\begin{eqnarray}
\label{th.int}
x_{0_1} & = & (104.57 \pm 0.11)\,{\rm cm} \\
x_{0_2} & = & (61.74 \pm 0.40)\,{\rm cm} \\ x_{0_3} & = & (37.56
\pm 0.31)\,{\rm cm}\,,
\end{eqnarray}
%\end{subequations}
with associated characteristic lengths
%\begin{subequations}
\begin{eqnarray}
\label{red.lengths}
l\left( x_{0_1}\right) & = & (121.44 \pm 0.09)\,{\rm cm}
\\ l\left( x_{0_2}\right) = l\left( x_{0_3}\right) = d & = &
(99.3 \pm 0.1)\,{\rm cm}\,.
\end{eqnarray}
%\end{subequations}
\begin{table}
\caption{\label{data}The experimental data. }
\begin{tabular}{lll}
 $x$\,(cm) & $T_1$\,(s) &
$T_2$\,(s) \\
\hline
 10 & 2.3613 & 2.0615 \\
 20 & 2.1492 & 2.0337 \\
 30 & 2.0363 & 2.0089 \\
 35 & 2.0016 & 1.9999 \\
 40 & 1.9838 & 1.9931 \\
 45 & 1.9733 & 1.9911 \\
 50 & 1.9754 & 1.9894 \\
 55 & 1.9799 & 1.9908 \\
 58 & 1.9846 & 1.9924 \\
 65 & 2.0055 & 2.0002 \\
 68 & 2.0173 & 2.0064 \\
 75 & 2.0470 & 2.0273 \\
 85 & 2.0939 & 2.0678 \\
 90 & 2.1224 & 2.0969 \\
 92 & 2.1334 & 2.1071 \\
 106 & 2.2178 & 2.2174 \\
 110 & 2.2441 & 2.2589 \\
 120 & 2.3078 & 2.3776 \\
\hline
\hline
\end{tabular}
\end{table}

Throughout the experiment the movable mass $m_{m}$ will be placed in
successive positions, generally 10\thinspace cm from each other, except
near
the theoretical characteristic positions (\ref{th.int}) where the distances
decrease (see the second column in Table~\ref{data})
\footnote [1]{
We
did not choose positions too close to the estimated characteristic
positions to prevent the casual occurrence of coincident period
measures about the two pivots.
In fact our distance measures are
effected by an error of $\approx \pm 1$\,mm. Such an error
would cause a strong distortion in determining the
empirical characteristic positions.
One of them would be directly determined by direct measure and
its error would not be lessened by the fitting procedure.
}
~.
\noindent The period of small oscillation about the two pivots are measured
for all those positions of $m_{m}$. These periods are measured by recording
the time of each of 9 consecutive oscillations when the pendulum starts
from
the angle $\varphi _{0}\sim 6^{\circ }\pm 1^{\circ }$. For this purpose we
used a photogate timed by an electronic digital counter
\footnote [2]{The resolution of the LEYBOLD-LH model is
0.1\,ms.}
~.
We
repeated the procedure for 18 positions of $m_{m}$, at first with respect
to
$c_{1}$ and then $c_{2}$. The average of the 9 values is taken to be the
period at the given position of $m_{m}$ whose error is given by half of its
maximum excursion,
that is, $\approx ~0.0018\,s$. The initial
angle $\varphi _{0}$ is sufficiently small that an equation similar to Eq.~
(%
\ref{eq.moto}) is valid. By expanding an elliptic integral in a power
series, it is possible to approximately express the period associated with
the exact equation of pendulum motion
\begin{equation}
\ddot{\varphi}+\frac{mgh_{i}}{I_{i}}\sin \varphi =0  \label{ex.eq.moto}
\end{equation}
by adding corrective terms,\cite{nelson,resnick} to the period expression
given in Eq.~(\ref{period}). In the next section we will evaluate such a
correction. The results are reported in Table~\ref{data}.

\section{The linear fitting procedure}

\label{sec4} We now describe a linear fitting procedure used to fit the
experimental data listed in Table~\ref{data}~ and empirically determine the
characteristic positions. The numerical computations were obtained using
MAPLE
\footnote[4] {We used Maple V, Release 5.1 by Waterloo Maple
Inc.}
 and some FORTRAN code
\footnote [5] {The FORTRAN
codes employed subroutines from Ref.~\cite{recipes} and the numerical
package NAG-Mark 14. The plotting package is PGPLOT 5.2 developed
by T. J. Pearson.}
~.
From a numerical
point of view we should fit the data by cubic polynomials like those in
Eq.~(%
\ref{equation:Ci}). Such a fitting can be treated linearly because the
coefficients $D_1$ and $D_2$ may be determined a priori by Eqs.~(\ref
{coeff1}%
) and (\ref{coeff2}) which involve only the known physical parameters
listed
in Table~\ref{parameters}. We obtain
%\begin{subequations}
\begin{eqnarray}
\label{d_i}
D_1 & = & (3.983\pm 0.01)\ 10^{-2}\,{\rm cm}^{-1}
\\ D_2 & = & (-4.2689 \pm 0.0047)\ 10^{-3}\,{\rm cm}^{-1}\,.
\end{eqnarray}
%\end{subequations}
We can obtain the desired fitting of the data obtained
in Sec.~\ref{sec3}
by
applying the least squares method to the following function
%inizio proposta
\begin{equation}
\Xi_{i} \left( A_i, B_i, C_i \right) =\sum_{h=1}^{18}\left(\frac{%
T_{h,i}^{2}- \frac {A_i x_h^2 + B_i x_h +C_i}{1 + D_i x_h}} {2
T_{h,i}\sigma_T}\right)^2\,,
\end{equation}
where $(x_h,T_{h,i})$ are the data of
the $i$th set in Table~\ref{data} 
\footnote [6] 
{We also know that $B_1=0$ and the number of
coefficients to be estimated by the fitting procedure can be
reduced. 
}~.

Two  sources of error
with  period measurements
were considered
\begin{itemize}
\item  
In any position and for both pivots, we  considered the standard
deviation  of the 9 period
electronic measurements,  
       varying from 0.0003\,s to 0.0036\,s .
\item  The systematic error that formula
       \ref{Tmodified}   introduces on the data. For example when
       T=2.3\,s (the maximum period here analysed)
       and   $\varphi _{0}\sim 12^{\circ }$ the shift introduced on
       T is  0.006\,s .
\end{itemize}
After this analysis we considered
$\sigma_T= 0.006$\,s  as estimated error for period measurements.
%fine proposta

\begin{figure}[htbp]
\begin{center}
\includegraphics[width=10cm,angle=-90]{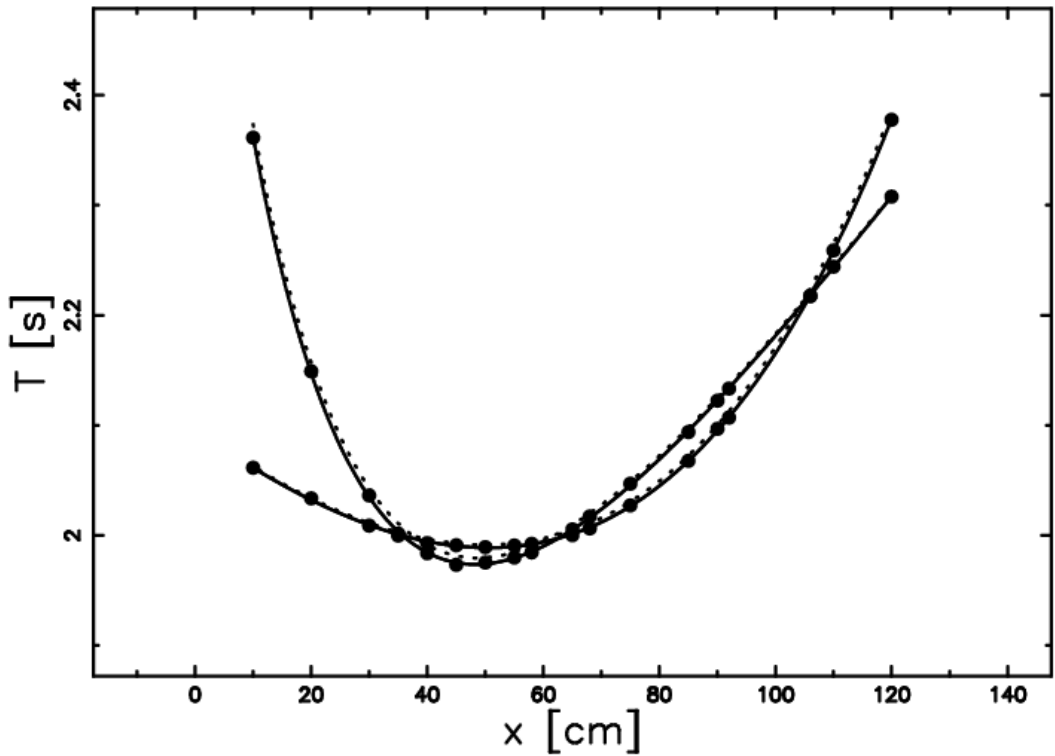}
\end{center}
\caption{\label{fig_04}Theoretical cubics (dotted line),
fitted cubics (full line).
and experimental data (filled points).
The experimental errors are much smaller than the filled points drawn,
so they are not visible within this plot.
}
\end{figure}

The obtained results are reported in
Table~\ref
{fitcoef1} ,  Table~\ref{fitcoef2}
and visualised in Fig.\ref{fig_04}.
\begin{table}
\caption{\label{fitcoef1}Coefficients of the cubic curve ${\mathcal C}_ 1$
estimated
by the
linear method.
}
\begin{tabular}{cc}
\hline
\hline
$A_1$   & $( 0.001607 \pm 0.000003) \,s^2 \,cm^{-2}$   \\
$B_1$   & $      0    \,               \,s^2   \,cm^{-1}$  \\
$C_1$\, & $(  7.641   \pm 0.011      )  \,s^2$ \\
\hline
\hline
\end{tabular}
\end{table}

\begin{table}
\caption{\label{fitcoef2}Coefficients of the cubic curve ${\mathcal C}_ 2$
estimated by
the linear method.}
\begin{tabular}{ll}
\hline
\hline
$A_2$   & ($0.000172 \pm 0.000002$)$ \,s^2 \,cm^{-2}$ \\
$B_2$   & ($-0.03422 \pm 0.00031$ )$ \,s^2 \,cm^{-1}$ \\
$C_2$   & ($4.393 \pm 0.01 $)      $ \,s^2  $ \\
\hline
\hline
\end{tabular}
\end{table}

%figure fig_04

The  merit function $\chi^2$ and
 the associated $p$--values are
reported in Table~\ref{p-values-linear} and each of them
 have  to be understood as the
maximum probability to obtain a better fitting.

\begin{table}
 \caption[]{$\chi^2$ and critical $p$--values for linear fitting by cubic
curves}
 \label{p-values-linear}
 \[
 \begin{array}{ccc}
 \hline
 \hline
 ~~~~ & {\mathcal C}_ 1, ~degrees~ of~freedom=16  &; {\mathcal C}_2 ~degrees~
~of~freedom=15\\
 \hline
 \noalign{\smallskip}

\chi^2  & 2.69 &  1.57  \\ \noalign{\smallskip}

\int_0^{\chi^2}\chi^2(x,15)dx  & 0.00008  &  0.000005 \\ \noalign{\smallskip}

\noalign{\smallskip}
 \hline
 \hline
 \end{array}
 \]
 \end {table}

The estimated cubic coefficients of Table~\ref{fitcoef1} and Table~\ref
{fitcoef2} allow us to evaluate the characteristic positions and the
associated characteristic periods by intersecting their upper
branches
\footnote [7] {
These two cubic curves represent the period-distance
relations in the plane $(x,T)$ when oscillations are considered about $c_1
$
or $c_2$ respectively. Then their common points coordinates give
the characteristic positions of the pendulum and the associated periods.
We have already observed in Sec.~\ref{sec2} that these cubics are
symmetrical
with respect to the $x$-axis. More precisely each of them is composed by
two
symmetrical branches. The branches lying under the $x$-axis are not
physically
interesting since their period coordinate $T$ is negative. Therefore the
only
interesting common points of these two cubic curves are the intersection
points
of their upper branches.
}
We obtain a cubic equation whose numerical solutions are
reported in Table~\ref{intersections}.

\begin{table}
\caption{\label{intersections}Estimated intersection points of
fitting cubic curves.}
\begin{tabular}{cc}
\hline
\hline
$(x_{0_1},T(x_{0_1}))$ & (106.015\,cm,\ 2.2184\,s) \\
$(x_{0_2},T(x_{0_2}))$ &(62.541\,cm,\ 1.9973\,s) \\
$(x_{0_3},T(x_{0_3}))$ &(35.779\,cm,\ 1.9998\,s) \\
\hline
\hline
\end{tabular}
\end{table}

 Refer to Eqs.~(\ref{g}) and (\ref
{red.lengths}) to compute the associated values of $g$. We have
%\begin{subequations}
\begin{eqnarray}
\label{g-fit}
g_1 = 4 \pi^2 \frac {l( x_{0_1})} {T(
x_{0_1})^2} \\ g_2 = 4 \pi^2 \frac {l(
x_{0_2})} {T( x_{0_2})^2} \\ g_3 = 4 \pi^2
\frac {l( x_{0_3})} {T( x_{0_3})^2}\,,
\end{eqnarray}
%\end{subequations}
and their numerical values are listed in Table~\ref{glinear}.

\begin{table}
\caption{\label{glinear}Values of $g$ obtained by formulas (\ref{g-fit})
and (\ref{datag}).}
\begin{tabular}{cc}
 \hline
 \hline
$g_1$          & ($974.15 \pm 2.72$) \,cm\,s$^{-2}$\\
$g_2$          & ($982.65 \pm 3.11$) \,cm\,s$^{-2}$\\
$g_3$          & ($980.20 \pm 3.1 $) \,cm\,s$^{-2}$\\
$\overline{g}$ & ($979.00 \pm 1.72$) \,cm\,s$^{-2}$ \\
\hline
\hline
\end{tabular}
\end{table}

Their average
gives
\begin{equation}
\overline{g} = (979.00 \pm 1.72)\, \mathrm{cm\,s}^{-2}\,.  \label{datag}
\end{equation}
%inizio proposta 
where the uncertainty  is  found implementing the error
propagation equation (often called law of errors of Gauss) when
the covariant terms are neglected (see  equation (3.14)
in~\cite{bevington}).
%fine proposta 
We now consider the correction arising from the
approximation of the exact
equation of pendulum motion (\ref{ex.eq.moto}) already mentioned at the end
of the previous section.
% inizio proposta
This correction gives:\cite{nelson,resnick}
\begin{equation}
T=
2\,\pi\,\sqrt {{\frac {l}{g}}} \left( 1+1/16\,{{\it \varphi _{0}}}^{2}
 \right)
\label{Tmodified}
\quad ,
\end{equation}
and
\begin{equation}
g=
4\pi ^{2}{\frac{l}{{T^{2}}}}
\left(
 1+1/16\,{{\it \varphi _{0}}}^{2}
\right )
^{2}
\,.  \label{gg}
\end{equation}
% fine proposta
A small increase in the value of $g$ is evident from Eq.~(\ref{gg}), and we
will refer to it as the finite amplitude correction (f.a.c.).

With the data listed in (\ref{glinear}) we obtain
\begin{equation}
\overline{g}_+ = (980.34 \pm 1.74)\,\mathrm{cm\,s}^{-2}\, ,
\end{equation}
which is the gravity acceleration increased by the f.a.c.. An accurate
measure of the value of $g$ in Turin\cite{gtorino}
gives
\begin{equation}
g_T =980.534099(4)\,\mathrm{cm\,s}^{-2} \,. \label{g_T}
\end{equation}
This value will be considered as the ``true value'' of the acceleration due
to the earth's apparent gravity field in Turin~
\footnote [8]
{
Further references for accurate measurements of $g$ are
the following.
A world-wide survey of all
the apparent gravity measurements (see {\tt {<}http://bgi.cnes.fr{>}})
gives for Turin
$g=980.5495\,cm\,s^{-2}$; this value differs from $g_T$ by
16\,ppm.
An analytical formula provided by the U.S.\ Geological
Survey~\cite{moreland}
  needs two input parameters,
the local height above sea level and latitude, which in our case
are 236\,m and $45.05333^{\circ}$ respectively to give the
local value of $g$. For Turin this formula gives
$g=980.5937$\,cm\,s$^{-2}$, which differs from $g_T$ by
61\,ppm.
}~.
By comparing
it with $\overline{g}_+$, we see that our measurement is $-191$\,ppm
smaller
than the "true value."

Note that the considered Kater pendulum admits characteristic positions
sufficiently distant from each other (see formulas (\ref{th.int}) and data
collected in Table~\ref{intersections}). Then it can be considered
sufficiently ``well--assembled'', which means that a parabolic fitting (of
type (\ref{parabola})), of the empirical
data $(x_{h},T_{h,i})$ collected in
Table~\ref{data}, should give a sufficiently
precise evaluation of
characteristic positions $x_{0_{2}},x_{0_{3}}$.
As before we     apply the least
square method to the following function ,
but the sum is now
extended to the first 13 entries
of  Table~\ref{data} in order to exclude the first intersection
\begin{equation}
\Theta _{i}\left( A_{i},B_{i},C_{i}\right) =
\sum_{h=1}^{13}\left( \frac{%
T_{h,i}-\left( A_{i}x_{h}^{2}+B_{i}x_{h}+C_{i}\right) }
{\sigma_{T}}\right) ^{2}\,.
\end{equation}
The coefficients of the fitting parabolas are reported in
Tables(~\ref{fitcoef1_p}) and
(~\ref{fitcoef2_p}) ;
the   $\chi^2$ and
 the associated $p$--values are
reported in Table~\ref{p-values_parabolic}
%inizio proposta
, in comparison the parabolic fit gives  very bad  results.
.
%fine   proposta
 \begin{table}
 \caption[]{Coefficients of  the first fitting parabola ${\mathcal P}_1$.}
 \label{fitcoef1_p}
 \[
 \begin{array}{cc}
 \hline
 \hline
 \noalign{\smallskip}

A_1          &  (0.000180 \pm 0.000002)\,s~cm^{-2}  \\ \noalign{\smallskip}

B_1          &  (-0.01959 \pm 0.00017)\,s~cm^{-1}  \\ \noalign{\smallskip}

C_1          &  (2.494 \pm 0.003)\,s               \\ \noalign{\smallskip}

\noalign{\smallskip}
 \hline
 \hline
 \end{array}
 \]
 \end {table}

\begin{table}
 \caption[]{Coefficients of  the second fitting parabola ${\mathcal P}_2$.}
 \label{fitcoef2_p}
 \[
 \begin{array}{cc}
 \hline
 \hline
 \noalign{\smallskip}

A_2  &  (0.000054 \pm 0.000002)\,s~cm^{-2} \\ \noalign{\smallskip}

B_2  &  (-0.00517 \pm 0.00017) \,s~cm^{-1} \\ \noalign{\smallskip}

C_2  &          (2.113 \pm 0.004)      \,s  \\ \noalign{\smallskip}

\noalign{\smallskip}
 \hline
 \hline
 \end{array}
 \]
 \end {table}

\begin{table}
 \caption[]{$\chi^2$ and critical $p$--values for linear fitting by
parabolas ,\\
first 13 data , degrees of freedom =10}
 \label{p-values_parabolic}
 \[
 \begin{array}{ccc}
 \hline
 \hline
 ~~~~ & {\mathcal P}_1 & {\mathcal P}_2 \\
 \hline
 \noalign{\smallskip}

\chi^2  & 893  &  12.5  \\ \noalign{\smallskip}

\int_0^{\chi^2}\chi^2(x,15)dx  & 1  & 0.74 \\ \noalign{\smallskip}

\noalign{\smallskip}
 \hline
 \hline
 \end{array}
 \]
 \end {table}

Their intersection points are given in
Table~(~\ref{intersections_p}).

\begin{table}
 \caption[]{Estimated intersection points of fitting parabolas. }
 \label{intersections_p}
 \[
 \begin{array}{cc}
 \hline
 \hline
 \noalign{\smallskip}

(x_{0_2},T(x_{0_2}))  &( 72.296 cm :  2.0207 s ) \\ \noalign{\smallskip}

(x_{0_3},T(x_{0_3}))  &( 41.709 cm :  1.9908 s ) \\ \noalign{\smallskip}

\noalign{\smallskip}
 \hline
 \hline
 \end{array}
 \]
 \end {table}

We get then the
following two evaluations of $g$:
%\begin{subequations}
\begin{eqnarray}
\label{g-fit_parabolic}
g' = 4 \pi^2 \frac {d} {T( x_{0_2})^2} = (960.08 \pm  3.00)\,\mathrm{cm\,s}
^{-2}\,    \\
g'' = 4 \pi^2 \frac {d} {T( x_{0_3})^2}= (989.11 \pm  3.14)\,\mathrm{cm\,s}
^{-2}\,\,.
\end{eqnarray}
%\end{subequations}
Their average gives
\begin{equation}
\overline{g}_{parabolic}=(974.60 \pm  2.18 )\,\mathrm{cm\,s}^{-2}\,.
\label{datag-parabolic}
\end{equation}
A comparison with $\overline{g}$ in (\ref{datag}) and $g_T$ in (\ref
{g_T}),
gives a clear evidence of the better efficiency of a cubic fit with
respect to a parabolic one.

\section{Conclusions}

We summarise  the main results of our theoretical and
numerical analysis
\begin {enumerate}
\item The three solutions of the Kater pendulum
      concerning the distance--period relationship
      discovered in 1907
      by Shedd and
      Birchby in 1907 \cite{shedd1, shedd2, shedd3}
      are classified in a modern context
\item The first  solution of  the distance--period relationship
      allows to deduce a new formula of $g$ via the
      second equivalent length both in the idealized
      pendulum and in a commercial Kater pendulum,
      see respectively formula(\ref{lidealized})
      and (\ref{l1})
\item One of the main targets of our work "the evaluation of g"
gives  oscillating results
\begin{itemize}
\item : our best numerical fit to $T^2$
produces  (the linear fit + non--linear correction) a value of
g that is  191 ppm smaller than the "true vale"
\item : our worst fit to $T^2$ (the non--linear fit +
non--linear correction)  gives a value of g that is
1978  ppm smaller than the "true vale"
\end {itemize}
\item  Concerning the fit to $T$ through a parabola
       we obtain high values
 of $\chi^2$ ($\chi^2$=893 for $C_1$ and $\chi^2$=12.51 for $C_2$)
 with respect  to the linear fit to  $T^2$
 ($\chi^2$=2.69  for $C_1$ and $\chi^2$=1.57  for $C_2$).
These high values of  $\chi^2$  allow to rule out the physical
significance of this type of fit.
\end {enumerate}

\ack{
We would like to thank G.Maniscalco for his assistance
in the preparation
of experiment set up.
We are also grateful to the anonymous referee for  useful
suggestions and improvements.
}

\appendix

\section{Reality of characteristic positions and structural conditions}

\label{appa} The characteristic positions of our pendulum are given by
Eqs.~(%
\ref{char.pos.1}) and (\ref{char.pos.2,3}). The former, $x_{0_{1}}$, is
always real. On the other hand $x_{0_{2}}$ and $x_{0_{3}}$ are real if and
only if the square roots in Eq. (\ref{char.pos.2,3}) are real i.e.e if and
only if
\begin{eqnarray}
m_{m}d^{2}+4mdK-4mK^{2}-4I_{0}^{\prime }\geq 0
\Longleftrightarrow   \nonumber \\
x_{f}^{2}-dx_{f}-\frac{(m_{m}+m_{b})d^{2}-4I_{0}^{\prime \prime }}{4m_{f}}%
\leq 0\,.  \label{cond}
\end{eqnarray}
The latter are the ``suitable conditions'' on the pendulum parameters of
Corollary \ref{PS}.

To avoid the overlapping of $m_f$ with $c_1$ we have to impose that
$x_f\leq
-r_f$. Then Eq.~(\ref{cond}) is equivalent to requiring that
\begin{equation}
m d^2 - 4 I_0^{\prime \prime}\geq 0 \ \mathrm{and}\ \frac{d}{2} - \frac{1}
{2}
\sqrt{\frac{m d^2 - 4 I_0^{\prime \prime}}{m_f}} \leq x_f \leq -r_f \,.
\label{struct.cond.3}
\end{equation}
Note that the condition on the right in Eq.~(\ref{struct.cond.3}) is not
empty if
\begin{eqnarray}
\frac{d}{2} - \frac{1}{2} \sqrt{\frac{m d^2 - 4 I_0^{\prime \prime}}{m_f}}
\leq -r_f \Longleftrightarrow  \nonumber \\
 d \geq 2\left(\frac{m_f r_f}{m_b + m_m}+
\sqrt{%
\left( \frac{m_f r_f}{m_b + m_m}\right)^2 + \frac{m_f r_f^2 + I_0^{\prime
\prime}}{m_b + m_m}}\right).  \label{mass cond.2}
\end{eqnarray}
In particular, the latter ensures that the left condition in Eq.~(\ref
{struct.cond.3}) is also satisfied because
\begin{equation}
m d^2 - 4 I_0^{\prime \prime} \geq 0 \Longleftrightarrow d \geq 2\sqrt
{\frac{%
I_0^{\prime \prime}}{m}} \,.
\end{equation}
To avoid the overlapping of $m_m$ with the knife--edges it follows that
either $r_m \leq x \leq d-r_m$ or $d+r_m \leq x \leq \frac{L+d}{2}$. After
some algebra we get the following results.

Assume that $m_m > m_f $ and set
%\begin{subequations}
\begin{eqnarray}
 M_1 & = & 2\frac{m_m r_m + m_f r_f}{m_m - m_f} \\ M_2 & = &
2\left( \frac{m_f r_f}{m_b + m_m} +
\sqrt{\left(\frac{m_f r_f}{m_b + m_m} \right)^2 +\frac{m_f r_f^2 +
I_0^{\prime \prime}}{m_b + m_m}} \right) \\ S_1 & = &
\frac{m_f-m_m}{2m_f} d + \frac{m_m}{m_f}r_m \\ S_2 & = &
\frac{d}{2} - \frac{1}{2} \sqrt{\frac{m d^2 - 4 I_0^{\prime
\prime}}{m_f}} \\ S_3 & = & \frac{d}{2} - \frac{1}{2}
\sqrt{\frac{m_f d^2 + m_b d^2 + 4m_m r_m d - 4 m_m r_m^2 - 4
I_0^{\prime \prime}}{m_f}}
\,,
\end{eqnarray}
%\end{subequations}
Then we have the following possibilities:

\noindent (1) $d < \min (M_1, M_2)$: in this case the pendulum admits only
one characteristic position given by $x_{0_1}$ because by Eqs.~(\ref
{struct.cond.3}) and (\ref{mass cond.2}), $x_{0_2}$ and $x_{0_3}$ are not
real; $x_{0_1}$ is not between the knives, but occurs on the opposite side
of the bar with respect to $m_f$; the system is in an almost symmetrical
mass configuration of the pendulum.

\noindent (2) For $\min (M_1, M_2)\leq d < \max (M_1, M_2)$, we have the
following possibilities:

(2a) If $M_1 < M_2$: we get only the characteristic position $x_{0_1}$
which
is between the knives if and only if $S_1 \leq x_f \leq -r_f$; otherwise, $
x_{0_1}$ is placed like in (1).

(2b) If $M_{2}<M_{1}$, we get all the characteristic positions; $x_{0_{1}}$
is like in (1) and $x_{0_{2}},x_{0_{3}}$ occur between the knives if and
only if $S_{2}\leq x_{f}\leq \min (S_{3},-r_{f})$.

\noindent (3) For $\max (M_1, M_2)\leq d$, the pendulum admits all the
characteristic positions $x_{0_1},x_{0_{2}},x_{0_{3}}$ which are placed as
follows:

(3a) Only $x_{0_1}$ is placed between the knives when either
\begin{equation}
S_1 < S_2\ \mbox{and}\ S_1 \leq x_f \leq S_2 \,,
\end{equation}
or
\begin{equation}
S_3 < -r_f \ \mbox{and}\ \max \left(S_1, S_3\right) < x_f \leq -r_f \,.
\end{equation}
(In particular, if $S_3 < S_1$, we can also assume the position $x_f = S_1$
for the fixed weight $m_f$.)

(3b) only $x_{0_{2}},x_{0_{3}}$ are placed between the knives when $S_2 <
S_1 $ and $S_2 \leq x_f < S_1$.

(3c) We obtain all the possible characteristic positions $%
x_{0_1},x_{0_{2}},x_{0_{3}}$ between the knives when
\begin{equation}
\max (S_1, S_2)\leq x_f \leq \min (S_3, -r_f)\,.
\end{equation}

In the concrete case considered in the Section~\ref{sec3} we have
%\begin{subequations}
\begin{eqnarray}
M_1 & = & (62.61 \pm 0.25)\,{\rm cm} \\ M_2 & = &
(70.87 \pm 0.05)\,{\rm cm}\\ S_1 & = & (-12.28 \pm 0.08)\,{\rm cm}
\\ S_2 & = & (-28.049 \pm 0.075)\,{\rm cm} \\ S_3 & = &
(-7.62 \pm 0.06)\,{\rm cm} \,,
\end{eqnarray}
%\end{subequations}
%inizio proposta 
where the uncertainty is found by applying the law of errors of
Gauss with the uncertainties listed in Table~\ref{parameters}.
%fine proposta 
Therefore we are in the case 3.b.

\section  {Further numerical methods}

\label{appc} We outline three additional numerical methods that may be
applied to analyse experimental data. The final results obtained by means
of
each method are reported in Table~\ref{ppm}.

\begin{table}
\caption[]{\label{ppm}Average values $\overline{g}$ and corrected values
$\overline{g}_+$
(by f.a.c.).}
\begin{tabular}{|l|c|c|}
\hline
\hline
algorithm &
$\overline{g}$     & $\overline{g}_+$  \\
\hline
linear fitting by parabolas & ($ 974.6 \pm  2.17$)\,$cm\,s^{-2}$
& ($  975.93 \pm  2.20$)\,$cm\,s^{-2}$ \\
linear fitting by cubics & ($979.00 \pm 1.72$)\,$cm\,s^{-2}$
& ($980.34 \pm 1.74$)\,$cm\,s^{-2}$ \\
non-linear fit & ($977.25 \pm 1.71$)\,$cm\,s^{-2}$
& ($978.25 \pm 1.74$)\,$cm\,s^{-2}$ \\
Cramer interpolation  & ($980.06 \pm 4.88$)\,$cm\,s^{-2}$
& ($981.40 \pm 4.89$)\,$cm\,s^{-2}$ \\
Spline interpolation  & ($979.52 \pm 1.73$)\,$cm\,s^{-2}$
&($980.86 \pm 1.74$)\,$cm\,s^{-2}$ \\
\hline
\hline
\end{tabular}
 \end{table}

\subsection{The non-linear method}

In the fitting procedure of data reported in Table \ref{data}, all the
coefficients $A_i,B_i,C_i$ and $D_i$ are considered as unknown parameters
to
be estimated. Therefore a fitting procedure performed by means of cubic
polynomials like those in Eq.~(\ref{equation:Ci}) is necessarily a
non-linear one. We want to apply the least square method to minimise the
following functions
\begin{equation}
X_i \left( A_i, B_i, C_i,D_i \right)= \sum_{h=1}^{18} \left ( T_{h,i}^2 -
{%
\frac {A_i x_h^2 + B_i x_h + C_i} {1+D_i x_h}} \right),  \label{residuals}
\end{equation}
which are non-linear in the unknown coefficients. The procedure is to apply
the NAG-Mark14 subroutine E04FDF to find an unconstrained minimum of a sum
of 18 nonlinear functions in 4 variables
(see Ref.~\cite{nag}).

The final value of $g$ is reported in the Table~\ref{ppm}.

\subsection{The Cramer interpolation method}

We present here a method that reduces our analysis in a local neighbourhood
of the estimated characteristic positions where a cubic behaviour of the
fitting curves is imposed.

{}From Eqs.~(\ref{coeff1}), (\ref{coeff2}), and (\ref{d_i}) we know that $%
D_1 $ and $D_2$ are completely determined by the pendulum parameters.
Moreover, from Eqs.~(\ref{coeff1}) we know that $B_1=0$. So to recover the
remaining coefficients of ${\mathcal C}_1$ and ${\mathcal C}_2$, we need to
interpolate two points of the first set of data in Table~\ref{data} and
three points of the second one respectively. We have to solve a $2\times 2$
and a $3\times 3$ linear system by applying the Cramer theorem (which is
the
most practical method for solving a linear system of equations). If we
choose data points that are close to a characteristic position, then the
nearest point in ${\mathcal C}_1 \cap { \mathcal C}_2$ to the chosen data will
give an empirical estimation of such a characteristic position and its
associated period. An iterated application of this procedure will produce a
distribution of periods and we may obtain $g$ from the mean value and its
statistical error from the standard deviation (see the last line of Table~%
\ref{cramer_data_fortran}).

\begin{table}
\caption{\label{cramer_data_fortran}The Cramer interpolation method.}
\begin{tabular}{|l|l|l|l|l|l|l|}
\hline
\hline
\multicolumn {7}{|c|} {Cramer Method} \\ \hline
Char. & \multicolumn {2}{c|} {Chosen data} & \multicolumn {2}{c|}
{Intersection} & Char. & g \\
 position & Series 1 & Series 2 & Position  & Period  & length
 & ~~ \\ \hline
$x_{0_1}$ & 15;18 & 15;17;18 & 105.773 \,cm  & t1,1= 2.217 \,s & 121.44 \,cm
& 975.73 \,cm\,s$^{-2}$\\
$x_{0_1}$ & 15;17 & 15;16;17 & 106.360 \,cm& t1,2= 2.221 \,s & 121.44\,cm
& 971.92 \,cm\,s$^{-2}$\\
$x_{0_1}$ & 16;18 & 16;17;18 & 106.108 \,cm& t1,3= 2.218 \,s& 121.44\,cm
& 974.08 \,cm\,s$^{-2}$\\
$x_{0_1}$ & 15;18 & 15;16;18 & 106.189 \,cm& t1,4= 2.219 \,s& 121.44\,cm
& 973.44 \,cm\,s$^{-2}$\\
$x_{0_2}$ & 7; 9 & 7; 8; 9 & 62.789 \,cm& t2,1= 1.996 \,s& 99.30\,cm
& 983.77 \,cm\,s$^{-2}$\\
$x_{0_2}$ & 8;10 & 8; 9;10 & 62.056 \,cm& t2,2= 1.996 \,s& 99.30\,cm
& 983.79 \,cm\,s$^{-2}$\\
$x_{0_2}$ & 9;11 & 9;10;11 & 61.962 \,cm& t2,3= 1.996 \,s& 99.30\,cm
& 984.28 \,cm\,s$^{-2}$\\
$x_{0_2}$ & 10;12 & 10;11;12 & 61.990 \,cm& t2,4= 1.996 \,s& 99.30\,cm
& 984.44 \,cm\,s$^{-2}$\\
$x_{0_3}$ & 2; 4 & 2; 3; 4 & 35.477\,cm & t3,1= 1.999 \,s & 99.30\,cm
& 980.87 \,cm\,s$^{-2}$\\
$x_{0_3}$ & 3; 5 & 3; 4; 5 & 36.207\,cm & t3,2= 1.998 \,s& 99.30\,cm
& 981.97 \,cm\,s$^{-2}$\\
$x_{0_3}$ & 4; 6 & 4; 5; 6 & 35.557\,cm & t3,3= 1.999 \,s& 99.30\,cm
& 981.11 \,cm\,s$^{-2}$\\
$x_{0_3}$ & 5; 7 & 5; 6; 7 & 36.668\,cm & t3,4= 1.995 \,s& 99.30\,cm
& 985.38 \,cm\,s$^{-2}$\\
\hline
\multicolumn{7}{|c|} {$\overline{g}$=( 980.06 $\pm$ 4.89)\,cm\,s$^{-2}$} \\
\hline
\hline
\hline
\end{tabular}
\end{table}

The results obtained for every interpolation are reported in Table~\ref
{cramer_data_fortran}; the chosen data points in the second and third
columns are enumerated as they appear in Table~\ref{data}.

\subsection{Cubic Spline Interpolation}

The last data analysis method to be proposed is the cubic spline
interpolation (subroutine SPLINE and SPLINT from Numerical Recipes~II).
Once
the three intersections are obtained, the procedure was similar to the
linear/nonlinear case and the final value of $g$ is reported in Table~\ref
{ppm}.

\end{document}